\documentclass[showpacs,showkeys,preprintnumbers,amsmath,amssymb,nofootinbib]{revtex4}

\usepackage{graphicx}
\usepackage{dcolumn}
\usepackage{bm}

\providecommand{\beqa}{\begin{eqnarray}}
 \providecommand{\bf}{\mathbf}
 \providecommand{\rm}{\mathrm}
\providecommand{\eeqa}{\end{eqnarray}}

\def\Z2{{\mathbf{Z}_2}}

\def\one{{\hbox{1\kern-.8mm l}}}
\newcommand{\beq}{\begin{equation}}
\newcommand{\eeq}{\end{equation}}
\def\be#1\ee{\begin{align}#1\end{align}}
\newcommand{\ov } {\over }

\def\vk{{\vec k}}

\def\ein{{\eta_{\text{in}}}}

\def\e{{\epsilon}}
\begin{document}
\preprint{\rightline{UUITP-12/11 }}
\title{Effects of Nonlinear Dispersion Relations on Non-Gaussianities}
\author{Amjad Ashoorioon$^{a}$}
\email{amjad.ashoorioon@fysast.uu.se}
\author{Diego Chialva$^{b}$}
\email{diego.chialva@umons.ac.be}
\author{Ulf Danielsson$^{a}$}
\email{ulf.danielsson@physics.uu.se}
\affiliation{$^a$Institutionen f\"{o}r fysik och astronomi
Uppsala Universitet, Box 803, SE-751 08 Uppsala, Sweden}
\affiliation{$^b$Universit\'e de Mons, Service de Mecanique et gravitation, Place
du Parc 20, 7000 Mons, Belgium}
\date{\today}


\begin{abstract}
We investigate the effect of non-linear dispersion relations on the
bispectrum. In particular, we study the case were the modified
relations do not violate the WKB condition at early times, focusing on a
particular example which is exactly solvable: the Jacobson-Corley
dispersion relation with quartic correction with positive
coefficient to the squared linear
relation. We find that the corrections to the
standard result for the bispectrum are suppressed by a factor
${H^2 \ov p_c^2}$ where $p_c$ is the scale where the modification to the
dispersion relation becomes relevant. The modification is {\it mildly}
configuration-dependent and equilateral configurations are more
suppressed with respect to the local ones, by a factor of one
percent. There is no configuration
leading to enhancements. We then analyze the results in the framework of particle
creation using the approximate gluing method of
Brandenberger and Martin, which relates more directly to the modeling
of the trans-Planckian physics via modifications of the
vacuum at a certain cutoff scale. We show that the gluing method
overestimates the leading order correction to the spectrum and
bispectrum by one and two orders, respectively, in ${H \ov p_c}$. We
discuss the various
approximation
and conclude that for dispersion relations not violating WKB at
early times the particle creation is small and does not lead to
enhanced contributions to the
bispectrum. We also show that in many cases enhancements do not occur
when modeling the trans-Planckian physics via modifications of the
vacuum at a certain cutoff scale. Most notably they are only of order
$O(1)$ when the Bogolyubov coefficients accounting for particle creation are determined by
the Wronskian condition and the minimization of the uncertainty
between the field and its conjugate momentum.

\end{abstract}




\maketitle

\section{Introduction}

The interest in non-Gaussian features of density perturbations has
been mounting in recent years. It is prompted by the upcoming release of the
data of unprecedentedly accurate experiments such as Planck and by more
advanced research activities. In particular, a notable interest descends from the possible
sensitivity of higher-point amplitudes on high-energy physics, namely
the existence of new physics at high energy scales. This is due to the fact
that higher correlation functions are suggested to be more sensitive to such
high energy effects than the power spectrum itself.

Non-Gaussianity is the non-zero value of higher-points ($>2$)
functions \footnote{The two-point one is related to the power spectrum.} of the comoving
curvature perturbation $\zeta$. In particular, here
we will be interested in the {\em bispectrum}, which is obtained
from the three-point function of the curvature perturbation. The
contributions to this correlator arise from different type of
diagrams depending on how we define the curvature perturbation (that
is, if we use non-linear field redefinitions). In particular we are
interested in the connected three-point function and in the effects
that modification of the high-energy theory can impart to the standard
slow-roll result. The outcomes of this analysis are important in two
respects: for the observation of such effects but also for the
potential risk for perturbation theory, which might break down
(trans-Planckian issue).

The three-point function for the conventional single-field slow-roll
models of inflation was originally calculated in \cite{Maldacena:2002vr}
and found to be slow-roll suppressed. Any non-Gaussian
signature is therefore a smoking-gun for deviation from the orthodox
picture. The effect one would generally expect are two-fold:

\begin{itemize}

 \item a peculiar shape function\footnote{That is a peculiar
 dependence of the result on the external momenta, leading to a
particular shape for the graph of this function.}

 \item an enhancement factor for some specific configurations.

\end{itemize}

Generally these effects are combined, and we have particular
enhancement factors when the external momenta of the three point
function (in momentum space) have specific
configurations.

Various methods have been employed to model high
energy modifications to the standard theory.
For example, the effects of the trans-Planckian physics have been implemented via
a stringy-modified space-time uncertainty relation \cite{Easther:2001fi}, or utilizing
the concept of a boundary action, which provides the boundary
conditions for the perturbations fields, subject to the renormalization from higher
energy scales \cite{Schalm:2004qk,Greene:2004np}. A particularly interesting method
is related to the choice of boundary conditions (the so-called
``choice of vacuum'') for the solution of the
equation of motion of the Mukhanov variable at the new physics
hyper-surface (NPHS), which corresponds to the time
the physical momentum of
perturbations reaches the scale $\Lambda$
\cite{Danielsson:2002kx,Danielsson:2002qh,Bozza:2003pr}. In this case,
the power
spectrum exhibits superimposed oscillations whose amplitude is
given by $H/\Lambda$, where $H$ is the Hubble parameter during the
inflation. The bispectrum is instead
found to be modified by enhancement factors for the enfolded/flattened
configuration of momenta, whose real magnitude
depends however on the value of the Bogolyubov
parameters calculated within this approach
\cite{Chen:2006nt,Holman:2007na, Meerburg:2009fi}. In such a
configuration, two of the momenta are
collinear with the third one in the momenta triangle.

In an alternative approach to model the effects of trans-Planckian
physics, that we will mainly focus on here, the
equation of motion for the perturbations are modified and new
dispersion relations, which differ from
the linear one for physical momenta larger than a fixed scale of new
physics $p_c$,
are considered \cite{Martin:2000xs}. Such {\em modified dispersion
relations} represent the violation of Lorentz invariance and were
also employed to analyze the possible effect of trans-Planckian
physics on black hole radiation \cite{Corley:1996ar}.
Modified dispersion relations of the kind we will investigate here, can
also be derived naturally from the recently proposed
Ho\v{r}ava-Lifshitz gravity
\cite{Horava:2009uw,Horava:2009if,Koh:2009cy} which is a
non-relativistic renormalizable modification of the gravity in the UV
region. They also arise in effective theory of single field inflation when the scalar perturbations propagate with a small sound speed \cite{Baumann:2011su}.

In this paper we focus on examples where one can obtain the exact
solutions to the field equations and therefore has control over
subtle effects. As the evolution of the mode function after the
modified dispersion relation has entered the linear regime can be mapped to an
excited state in the NPHS approach, using the exact solutions we have a
precise understanding and quantification of
particle creation and can understand the reason that lies behind the
presence or absence of the enhancement factors in the bispectrum.
The general result that we find is that, in absence of violation of
the WKB condition at early times, the enhancement factors are not
present in the bispectrum. On the other hand one would expect a more
favorable situation for dispersion relations with early time WKB
violation. However, as we have found no example of such modified
dispersion relations where the field equation was exactly solvable, we
discuss this case only briefly and leave the analysis of this point to
future research.

We also show how the enhancement factors are at maximum of order
$O(1)$ in some of the emergent
approaches to trans-Planckian physics in the NPHS framework.
Notably, this occurs in
the case where the Bogolyubov coefficients accounting for particle
creation are determined by
the minimization of the uncertainty relation
between the field and its conjugate momentum  and by the Wronskian
condition \cite{Danielsson:2002kx}, which yields the largest correction
to the spectrum.

We will first study a minimal modification to the linear dispersion
relation, known as Jacobson-Corley (JC)
\cite{Corley:1996ar}, both solving the field equations exactly, and
via approximation methods.
The JC  dispersion relation has a positive quartic correction to
the linear term and there is no WKB violation at
early times\footnote{The only WKB violation occurs when the modes exits
the horizon at late times, when the dispersion relation has become
approximately the standard linear one.}.

We find no modulation in the modified spectrum of perturbations, and
the amplitude of perturbations is damped quadratically with
the increase of the ratio ${H \ov p_c}$. This is
explained in terms of the very small particle creation, due to the
absence of WKB violation at early times.

We then turn to the analysis of the modification to the bispectrum:
using the exact solution to compute the three-point function,  we find no large
enhancement factors, and no interference terms as the one studied in
\cite{Chen:2006nt,Holman:2007na,Meerburg:2009fi}.
We then try to understand and interpret these
results in the language of particle creation using the gluing
approximation of Brandenberger and Martin
\cite{Martin:2000xs} for the solutions. We show that the gluing method overpredicts the magnitude of modification.

The outline of the paper is as follows: we review the formalism,
the notation and the techniques to solve the field equations in
section \ref{formalismnotation}. We then study
the JC modified dispersion relation in
section \ref{studyJCdispersion}: first we solve exactly the field
equations in section \ref{solvingJC}, then we study the two- and three-point functions
in sections \ref{secspectrumJC} and \ref{secbispectrumJC}. We then try to explain
and interpret the results in the framework of particle creation, and
make contact with the NPHS
framework in section \ref{secJCinterpretation}. We finally discuss the
general results, present the outlooks regarding modified dispersion relations
with early time violation of WKB condition, and conclude in
section \ref{discussionconclusion}.


\section{Formalism and notation.}\label{formalismnotation}


The starting point for our analysis is the definition of the curvature
perturbation of the comoving hypersurface
 \beq
 \zeta(\eta, x) = \int {d^3 k \over (2\pi)^{{3 \over 2}}} \zeta_{\vk}(\eta)
 e^{i \vec{k} \cdot \vec{x}},
 \eeq
where $\zeta_{\vk}(\eta)$ is defined in terms of the mode function $u_{\vk}(\eta)$ as
 \beq
  \zeta_{\vk}(\eta) = {u_{\vk}(\eta) \ov  z},
 \eeq
with equation of motion
\beq \label{eqofmo}
u_\vk'' + \left(\omega(\eta,\vk)^2- {z'' \over z} \right) u_\vk = 0
\eeq
where\footnote{$\phi$ being the inflaton field.} $z = {a\dot \phi \ov
H} $ and $\omega(\eta,\vk)$ is the comoving frequency as read from the
effective action. In the standard formalism $\omega(\eta,\vk)$ is taken
to be equal to $k\equiv |\vk|$. Instead, a modified dispersion
relation would correspond to considering different dependence of
$\omega$ on $\vk, \eta$. For isotropic backgrounds, which we will
confine ourselves to, one expects $u_{\vk}(\eta)$ and
$\omega(\eta,\vk)$ to depend only on $k$. Hence, we may drop the arrow
symbol at some points.

The three-point function is given by the formula
\beq \label{threepoint}
\langle \zeta(x_1)\zeta(x_2)\zeta(x_3)\rangle =
-2 \text{Re}\left( \int^\eta_\ein d \eta' i
\langle\psi_{\text{in}}|\zeta(x_1)\zeta(x_2)\zeta(x_3)
H_I(\eta')|\psi_{\text{in}}\rangle\right)
\eeq
where $H_I=-\int d^3 x a^3({\dot\phi \ov H})^4 H \zeta^{' 2}
\partial^{-2}\zeta'$ is the interaction Hamiltonian as defined in
\cite{Maldacena:2002vr}. $\ein$
and $|\psi_{\text{in}}\rangle$ are the initial time and state (vacuum). The
standard result in slow-roll inflation is obtained with $\ein = -\infty,$
and $|\psi_{\text{in}}\rangle$ is taken to be the Bunch-Davies vacuum.

In momentum phase space the three-point function becomes \cite{Maldacena:2002vr}
\be \label{bispectrum}
\langle \zeta_{\vk_1}(\eta)\zeta_{\vk_2}(\eta)\zeta_{\vk_3}(\eta)\rangle &=
2 \text{Re} \left(-i (2\pi)^3 \delta(\sum_i \vk_i)({\dot \phi \ov H})^4
{H \ov M_{P}^2} \int^\eta_{\eta_{\text{in}}} d\eta' {{a(\eta')^3} \ov k_3^2} \prod_{i=1}^3
\partial_{\eta'}G_{\vk_i}(\eta, \eta')
+ \text{permutations}\right)
\ee
where $\eta$ is taken to be a late time when all three functions $\zeta_{\vk_i}$
are outside the horizon\footnote{In the case of modified dispersion
relations, at this time the relation has become effectively the
standard linear one.}.
$G_{\vk_i}(\eta, \eta')$ is the Whightman function defined as:
\begin{equation}\label{Wightman}
G_{\vk}(\eta,\eta')\equiv\frac{H^2}{{\dot{\phi}}^2} \frac{u_k(\eta)}{a(\eta)}\frac{u_k^{\ast}(\eta')}{a(\eta')}
\end{equation}

We define the bispectrum following \cite{Babich:2004gb}
\beq\label{shape-func}
\langle
\zeta_{\vk_1}(\eta)\zeta_{\vk_2}(\eta)\zeta_{\vk_3}(\eta)\rangle
=\delta(\sum_i \vk_i) (2\pi)^3 F(\vk_1,\vk_2,\vk_3, \eta),
\eeq
where the translational invariance has imposed the conservation of
momentum. Scale-invariance requires that the function, $F$, to be a
homogeneous function of degree $-6$ and the rotational invariance
imposes it to be only a function of two variables, say $x_2\equiv
 k_2/k_1$ and $x_3\equiv k_3/k_1$. To avoid counting the same
configuration twice, it is further assumed that $x_3\leq x_2$. The
inequality $1-x_2\leq x_3$ also comes from the triangle
inequality.

The function $F$ contains a lot of information
about the source of non-Gaussianity and could be used to distinguish
among different models. The limit in which the configuration is such that
$x_3\simeq 0$ and $x_2\simeq 1$ is recognized in the literature as
the local configuration. The one in which $x_2\simeq x_3\simeq 1$ is
known as equilateral one.

Inflationary models in which non-linearity is developed
beyond the horizon, tend to produce a more local type of
non-Gaussianity. On the other hand, for the ones that the correlation
is among the modes with comparable wavelength, equilateral type of
non-Gaussianities tend to be produced \cite{Babich:2004gb}. These
modes will exit the horizon around the same time.

\subsection{Solving the field equation.}\label{solvingequations}

In the case of a modified
dispersion relation, equation (\ref{eqofmo}) is often solved
by approximation \cite{Martin:2000xs}. Two methods have been used: one
is the WKB approximation. This approach is convenient, as it is independent of
small parameters in the dispersion relation, because it only requires
a slow variation of $U^2 = \omega^2 -{a''\ov a}$ in time, {\it i.e.}
\cite{Martin:2002vn}.

 \beq \label{adiabaticity}
\left|{Q \ov U^2}\right| = \left|-{U'' \ov 2 U^3}+ 3{U^{'2} \ov 4 U^4}\right| \ll 1
\eeq
The other is the gluing procedure of Brandenberger and Martin, where
approximated solutions valid in different regions of behavior of
$\omega$ are glued together asking for the continuity of the functions
and their first derivatives \cite{Martin:2000xs}.

In the NPHS approach, enhancement terms appear in the bispectrum due to the
interference terms between negative and positive frequency parts of
the Wightman function as one starts from a non-Bunch-Davies vacuum. In
the case of modified dispersion relations, similar interference term
will generally arise solving (\ref{eqofmo})
using the gluing procedure. However, in all cases the actual
presence/absence of enhancement depends
on the magnitude of the second Bogolyubov parameter $\beta_{\vec k}$
\cite{Chen:2006nt,Holman:2007na,Meerburg:2009fi}.

With the knowledge of the exact solution in some specific examples, one can gain more precise
information about the shape of the bispectrum.
We want to explore these features in the case of
modified dispersion relations and in particular verify the presence/absence of
enhancement factors and their difference with the NPHS case where one employs the cut-off.
To do this, in the following we will consider a particular (but very illustrative)
example where the exact solution can be found. We then analyze and interpret the results in the framework of particle creation with
the approximation methods described above, and discuss the differences.


\section{Corley-Jacobson dispersion relation with $\mathrm{b_1>0}$. No
violation of WKB approximation.}\label{studyJCdispersion}


In this section, we will focus on Jacobson-Corley dispersion relation
with positive quartic correction. We assume that at very high physical
momenta, $p\gtrsim p_c$, the linear dispersion relation $\omega=p$
gets modified as follows:
\begin{equation}\label{JC}
\omega_{\text{phys}}^2=F^2(p)=p^2+\frac{b_1 p^4}{p_c^2}
\end{equation}
The dispersion relation is motivated by studies in black-hole physics
\cite{Corley:1996ar}, and was later used to study the effect of
trans-Planckian physics on cosmological perturbations
\cite{Martin:2000xs,Martin:2002kt}. Here, we will find an exact
analytic form of the solution and
will derive the effect of this modified dispersion relation on the two
and three point functions analytically. Then, we study the same
problem making use of approximation techniques and discuss the results.

\subsection{Exact solution of the field equations}\label{solvingJC}

In usual quantum field theory, the cosmological perturbations satisfy
eq.\eqref{eqofmo} as the equation of motion. A modified dispersion
relation of the kind of Corley-Jacobson entails the replacement
\begin{equation}\label{impl-CJD}
k^2\Rightarrow a^2(\eta) F^2\left(\frac{k}{a(\eta)}\right)=k^2+b_1 \frac{k^4}{p_c^2 a^2(\eta)}
\end{equation}
Then in de-Sitter space-time, where $a=-\frac{1}{H\eta}$,  equation
(\ref{eqofmo}) reads:
\begin{equation}\label{eom-cosm-perturb}
u''_{k}(\eta)+(k^2+\e^2 k^4 \eta^2-\frac{2}{\eta^2}) u_k=0
\end{equation}
where,
\begin{equation}\label{e}
\e\equiv  \frac{\sqrt{b_1} H}{p_c}.
\end{equation}
The above differential equation has two independent closed-form exact
solutions that can be given in terms of ${\rm
WhittakerW}(a,b,z)$ and its conjugate ${\rm
WhittakerW}^{\ast}(a,b,z)$ functions. For brevity, we will show
these two functions as ${\rm WW}(a,b,z)$ and ${\rm WW}^{\ast}(a,b,z)$
from now on, respectively. The exact solution is then:
\begin{eqnarray}\label{WR-sol}
u_{k}&=&\frac{C_1}{\sqrt{-\eta}} {\rm WW}\left(\frac{i}{4\e},\frac{3}{4},-i \e k^2 \eta^2\right)+\frac{C_2}{\sqrt{-\eta}} {\rm WW}^{\ast}\left(\frac{i}{4\e},\frac{3}{4},-i \e k^2 \eta^2\right)\nonumber\\
&=&\frac{C_1}{\sqrt{-\eta}} {\rm WW}\left(\frac{i}{4\e},\frac{3}{4},-i \e k^2 \eta^2\right)+\frac{C_2}{\sqrt{-\eta}} {\rm WW}\left(\frac{-i}{4\e},\frac{3}{4},i \e k^2 \eta^2\right).
\end{eqnarray}
The above solution is subject to the Wronskian condition
\begin{equation}\label{Wronskian}
u(\eta) u'^{\ast}(\eta)-u^{\ast}(\eta) u'(\eta)=i,
\end{equation}
which impose the following constraint on $C_1$ and $C_2$:
\begin{equation}\label{wr-cons-c1c2}
2i \left|C_1\right|^2 \exp(\frac{\pi}{4\e}) \e k^2-2i \left|C_2\right|^2 \exp(\frac{\pi}{4\e}) \e k^2=i,
\end{equation}
This condition would not uniquely determine $C_1$ and $C_2$. One
has to make extra assumptions to determine $C_1$ and $C_2$. We
will assume that the mode function approaches the positive frequency
WKB at ealy times. As it was shown by
\cite{Martin:2000xs}, this  choice will minimize the energy density
too.

In the limit of $\eta\rightarrow -\infty$, the exact equation of motion is reduced to the following form
\begin{equation}\label{eom-approx}
u''_k(\eta)+\e^2 k^4 \eta^2 u(\eta)=0,
\end{equation}
and its approximate positive frequency WKB solution is:
\beq \label{WKB}
u_k(\eta)\simeq \frac{1}{\sqrt{2 \omega(\eta)}} \exp(-i\int^{\eta} \omega(\eta) d\eta')
= \frac{1}{k \sqrt{-2\e\eta}}\exp(\frac{i\e k^2 \eta^2}{2}).
\eeq
The limit of the exact solution (\ref{WR-sol}) for large time can be
found writing the Whittaker functions in terms of ${\rm KummerU}$ functions
and using the asymptotic form of the latter for large argument
\cite{Abramowitz}. We obtain\footnote{The extra factor of $(-i\e k^2
\eta^2)^{\frac{i}{4\e}}$ is the subleading correction to the WKB
approximation in the limit of $\eta\rightarrow -\infty$}
\begin{eqnarray}\label{asymptot-mode}
u_k(\eta) & = & \frac{C_1}{\sqrt{-\eta}} \exp(\frac{i \e k^2 \eta^2}{2}) (-i \e k^2 \eta^2)^{5/4} {\rm U}\left(\frac{i}{4\e},\frac{3}{4},-i \e k^2 \eta^2\right)\nonumber \\ &&+ \frac{C_2}{\sqrt{-\eta}} \exp(-\frac{i \e k^2 \eta^2}{2}) (i \e k^2 \eta^2)^{5/4} {\rm U}\left(\frac{-i}{4\e},\frac{3}{4},i \e k^2 \eta^2\right) \nonumber \\
&\simeq&\frac{C_1}{\sqrt{-\eta}} (-i\e k^2 \eta^2)^{\frac{i}{4\e}} \exp(\frac{i \e k^2 \eta^2}{2})
+ \frac{C_2}{\sqrt{-\eta}} (i\e k^2 \eta^2)^{-\frac{i}{4\e}} \exp(-\frac{i \e k^2 \eta^2}{2}),\nonumber \\
\end{eqnarray}
where we see that the positive
frequency WKB solution corresponds to the choice $C_2=0$. Using the Wronskian
condition, \eqref{wr-cons-c1c2}, $C_1$ is fixed up to a phase:
\begin{equation}\label{C2}
C_1=\frac{\exp(\frac{-\pi}{8\e})}{\sqrt{2\e} k},
\end{equation}
so that our exact complete solution to the mode equation is, finally,
\begin{equation}\label{exact-adiabatic}
u_{k}(\eta)=\frac{\exp(\frac{-\pi}{8 \e})}{\sqrt{-2\e\eta} k}  {\rm WW}\left(\frac{i}{4\e},\frac{3}{4},-i \e k^2 \eta^2\right).
\end{equation}

\subsection{Two-point function and power spectrum}\label{secspectrumJC}

\begin{figure}[t]
\includegraphics[angle=0,scale=0.5]{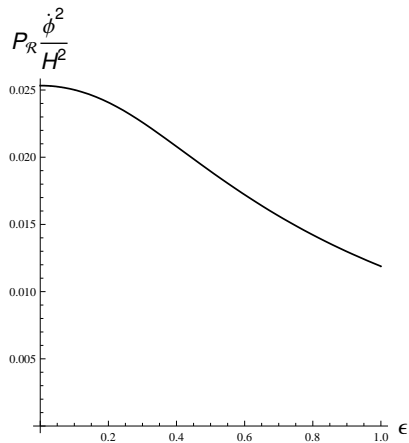}
\caption{The figure shows the dependence of the power spectrum in
de-Sitter space to the parameter $\e$, which quantifies the ratio of
Hubble parameter over the momentum scale $p_c$ around which the behavior
of the dispersion relation changes.} \label{power-plot}
\end{figure}

With the exact mode function in equation (\ref{asymptot-mode}) one calculates the power-spectrum
\begin{equation}\label{Pwr-Spectrum}
P_{\mathcal{R}}(k,\e)=\frac{k^3}{2\pi^2}\frac{H^2}{{\dot{\phi}}^2} \left|\frac{u_k(\eta)}{a(\eta)}\right|^{2}_{\frac{k}{aH}\rightarrow 0}.
\end{equation}
The spectrum is found to be still scale-invariant (no dependence on
$k$), but the standard result now shows corrections depending on $\e$:
\begin{equation}\label{Pwr-de-sitter}
P_{\mathcal{R}}(\e)=\frac{H^4}{\dot{\phi}^2}\frac{\exp(-\frac{\pi}{4\e})}{16 \pi \e^{3/2} \Gamma(\frac{5}{4}-\frac{i}{4\e})\Gamma(\frac{5}{4}+\frac{i}{4 \e})}.
\end{equation}
The plot of the power spectrum in terms of $\e$ is shown in figure
(\ref{power-plot}). The scale-invariance of the power spectrum is
again due to the scale-invariance of the de-Sitter
background. The corrections to the standard slow-roll are
\begin{equation}\label{Pwr-2nd}
P_{\mathcal{R}} (\e)\simeq \left(\frac{H^2}{2\pi\dot{\phi}}\right)^{2} \left(1-\frac{5}{4}\e^2\right).
\end{equation}
These corrections are appearing at second order in $\e$.

\subsection{Three-point Function}\label{secbispectrumJC}

Given the exact mode-function \eqref{exact-adiabatic}, the Wightman function for the positive frequency WKB
vacuum is
\begin{equation}\label{wightman-CJ}
G_{k}(\eta,\eta')=\frac{\exp(-\frac{\pi}{4 \e}) \sqrt{\eta\eta'}}{2 ek^2}\frac{H^4}{\dot{\phi}^2} {\rm WW}\left(\frac{i}{4\e}, \frac{3}{4}, -i \e k^2 \eta^2\right){\rm WW}\left(\frac{-i}{4\e}, \frac{3}{4}, i \e k^2 \eta'^2\right).
\end{equation}
To compute the three-point function from equation
(\ref{bispectrum}), the first argument of the
Wightman function has to be set at a time when the mode $k$ is outside
the horizon, generically taken to be time zero. Then, the Whightman
functions must be differentiated with
respect to the second argument. We obtain
\begin{eqnarray}\label{aGprime}
a(\eta) \partial_{\eta}G_{k}^{>}(0,\eta))&=& -\frac{(-1)^{\frac{1}{8}} H^3 \exp(-\frac{\pi}{4\e})\sqrt{\pi}}{8 \e^{\frac{9}{4}}k^{\frac{5}{2}}\dot{\phi}^2 (-\eta)^{\frac{3}{2}}\Gamma\left[\frac{5}{4}-\frac{i}{4\e}\right]} \left[-4\e {\rm WW}\left(1-\frac{i}{4\e}, \frac{3}{4}, i\e k^2 \eta^2\right) + (i+\e+2i\e^2k^2 \eta^2) \right. \nonumber \\&&\times\left.{\rm WW}\left(\frac{-i}{4\e}, \frac{3}{4}, i \e k^2 \eta^2\right)\right],
\end{eqnarray}
For computing the three-point function, it is convenient to write the
WhittakerW functions in terms of KummerM's based on the following relation
which is valid in the principal branch, $-\pi<{\rm arg}(z)\leq \pi$.
\begin{equation}\label{WW-M}
\mathrm{WW}(a,b,z)=\exp(-\frac{z}{2})\left(\frac{\Gamma(2b){\rm M}(\frac{1}{2}-a-b,1-2b,z)z^{\frac{1}{2}-b}}{\Gamma(\frac{1}{2}-a+b)}
+\frac{\Gamma(-2b){\rm M}(\frac{1}{2}-a+b,1+2b,z)z^{\frac{1}{2}+b}}{\Gamma(\frac{1}{2}-a-b)}
\right)
\end{equation}
We can then exploit the following expansion of the KummerM functions in terms of Bessel-$J$
functions \cite{Abramowitz}:
\begin{equation}\label{KummerM-BesselJ}
{\rm M}(a,b,z)=\Gamma(b) \exp(hz) \sum_{n=0}^{\infty} C_n z^n (-az)^{\frac{1}{2}(1-b-n)} J_{b-1+n} (2\sqrt{-az}).
\end{equation}
$C_{n+1}$ is given in terms $C_n,~C_{n-1}$ and $C_{n-2}$ from the following recursive relation
\begin{gather}\label{Cn}
C_{n+1}=\frac{1}{(n+1)}\left[((1-2h)n-bh)C_{n}+((1-2h)a-h(h-1)(b+n-1)) C_{n-1}-h(h-1)a C_{n-2}\right]. \\
C_{0}=1,\qquad C_1=-bh, \qquad C_2=-\frac{1}{2}(2h-1)a+\frac{b}{2}(b+1)h^2
\end{gather}
Although strictly valid for small $z$, the above expansion becomes the
correct expansion of the bispectrum for small $\e$ when the
integration over time in the formula (\ref{bispectrum}) for the
three-point function is performed. We will also prove this point in
section \ref{alternativeJCcomplete}.

As it appears from equations \eqref{aGprime} and (\ref{WW-M}), the function $a(\eta)
\partial_{\eta}G_{k}^{>}(0,\eta))$ is proportional to $\exp(\frac{-i
\e k^2 \eta^2}{2})$. Factoring out this term, one can expand the
coefficients in powers of $\e$ after the KummerM functions are
substituted with their approximate Bessel series expansion. To second
order in $\e$, we have:
\begin{equation}\label{G-approx}
a(\eta) \partial_{\eta}G_{k}^{>}(0,\eta)) \simeq \frac{H^3}{\dot{\phi}^2} \exp\left(\frac{-i \e k^2 \eta^2}{2}+ik\eta\right) \left[-\frac{1}{2k}-\e\frac{ ik\eta^2}{4}+\e^2 \left(\frac{5}{8k}-i\frac{5\eta}{8}-\frac{1}{8}k\eta^2-\frac{i}{12}k^2 \eta^3+\frac{1}{16}k^3 \eta^4\right)\right].
\end{equation}
It is now straightforward to obtain the three-point function. We will
perform the relevant integration over $\eta$ and expand the final
results in series of $\e$ in order to obtain the corrections to the
standard ($\e=0$) spectrum.
We show this procedure in detail for the zeroth order term in the square bracket
of (\ref{G-approx}), we then present the final results regarding the
other terms in the square brackets.

The zeroth order term in the square bracket of (\ref{G-approx}) leads to the contribution
\begin{eqnarray}\label{zeroth}
\langle \zeta_{\overrightarrow{k_1}}\zeta_{\overrightarrow{k_2}}\zeta_{\overrightarrow{k_3}}\rangle_{0}&=& -i (2\pi)^3 \delta^3\left(\sum \overrightarrow{k_i} \right) \left(\frac{\dot{\phi}}{H}\right)^4 M_P^{-2} H \int_{-\infty}^0 d\eta \frac{1}{k_3^2} \frac{-H^9}{8 \dot{\phi}^6 k_1 k_2 k_3} \exp\left[-i\eta (k_1+k_2+k_3)+\frac{i\e\eta^2}{2}(k_1^2+k_2^2+k_3^2)\right]\nonumber\\&+&{\rm c.c.}+{\rm permutations},
\end{eqnarray}
with solution
\begin{equation}\label{tp0}
\langle \zeta_{\overrightarrow{k_1}}\zeta_{\overrightarrow{k_2}}\zeta_{\overrightarrow{k_3}}\rangle_{0}=2 \Re\left[\frac{H^6}{\dot{\phi}^2 M_P^2}\frac{(-1)^{\frac{3}{4}} \pi^{\frac{7}{2}} \exp(-\frac{i k_t^2}{2 \e k_s^2}) {\rm Erfi}\left[\frac{(1-i) (-k_t+\e k_s^2 \eta)}{2\sqrt{\e k_s^2}}\right]}{k_1 k_2 k_3^3 \sqrt{2\e k_s^2}}\right]_{\eta=-\infty}^{\eta=0}+{\rm permutations},
\end{equation}
where $\Re$ indicates the real part and
\beq
k_t\equiv k_1+k_2+k_3 \,, \qquad k_s^2\equiv k_1^2+k_2^2+k_3^2 \, .
\eeq
In contrast to the
Lorentzian dispersion relation, $\omega=p$, the integrand remains
finite at $\eta=-\infty$ even {\it without} the change of variable
from $\eta\rightarrow\eta+i\epsilon|\eta|$. In this limit the ${\rm
Erfi}$ function tends to $i$. This is interesting, as the adhoc
prescription to make sense of the contribution at $-\infty$ is
completely taken care of by the modified dispersion relation, which is expected from quantum gravity.

Taking the real part and expanding to second order in $e$, one obtains
\begin{equation}\label{TP0e2}
\langle \zeta_{\overrightarrow{k_1}}\zeta_{\overrightarrow{k_2}}\zeta_{\overrightarrow{k_3}}\rangle_{0}\simeq \delta^{3}\left(\sum_i \overrightarrow{k_i}\right)\left[\frac{2H^6 \pi^3}{\dot{\phi}^2 k_1 k_2 k_3 k_t M_p^2 }-\e^2 \frac{6 H^6 k_s^4 \pi^3}{\dot{\phi}^2 k_1 k_2 k_3 k_t^5 M_p^2}\right] \left(\frac{1}{k_1^2}+\frac{1}{k_2^2}+\frac{1}{k_3^2}\right)
\end{equation}
The first term in the square brackets is the regular quantum field
theory result in absence of any nonlinearity in the dispersion
relation. The first correction is proportional to $\e^2$.

We then apply the same procedure of integration and expansion for
small $\e$ for the higher order terms in the square brackets of
(\ref{G-approx}). We obtain
\begin{itemize}
\item[-] for the term of order $\e$ in the square brackets
\begin{equation}\label{Tp1e2}
\langle \zeta_{\overrightarrow{k_1}}\zeta_{\overrightarrow{k_2}}\zeta_{\overrightarrow{k_3}}\rangle_{1}\simeq \delta^{3}\left(\sum_i \overrightarrow{k_i}\right) \e^2 \frac{12 H^6 k_s^4 \pi^3}{\dot{\phi}^2 k_1 k_2 k_3 k_t^5 M_p^2}\left(\frac{1}{k_1^2}+\frac{1}{k_2^2}+\frac{1}{k_3^2}\right),
\end{equation}
\item[-] for the term of order $\e^2$ in the square brackets
\begin{equation}\label{TP2e2}
\langle \zeta_{\overrightarrow{k_1}}\zeta_{\overrightarrow{k_2}}\zeta_{\overrightarrow{k_3}}\rangle_{2} \simeq \delta^3\left(\sum \overrightarrow{k_i} \right) \e^2 \frac{\pi^3 H^6 (6 k_s^4 - 2 k_c^3 k_t + k_s^2 k_t^2 + 10 k_t^4)}{ k_1 k_2 k_3 k_t^5 M_P^2 \dot{\phi}^2} \left(\frac{1}{k_1^2}+\frac{1}{k_2^2}+\frac{1}{k_3^2}\right),
\end{equation}
where
\beq
k_c^3=k_1^3+k_2^3+k_3^3 \, .
\eeq
\end{itemize}

Having all the contributions, we can compute the total three-point function:
\begin{equation}\label{TPtot}
\langle
\zeta_{\overrightarrow{k_1}}\zeta_{\overrightarrow{k_2}}\zeta_{\overrightarrow{k_3}}\rangle_{\rm tot}=
\langle \zeta_{\overrightarrow{k_1}}\zeta_{\overrightarrow{k_2}}\zeta_{\overrightarrow{k_3}}\rangle_{0}+
\langle\zeta_{\overrightarrow{k_1}}\zeta_{\overrightarrow{k_2}}\zeta_{\overrightarrow{k_3}}\rangle_{1}+
\langle \zeta_{\overrightarrow{k_1}}\zeta_{\overrightarrow{k_2}}\zeta_{\overrightarrow{k_3}}\rangle_{2},
\end{equation}
and the relative change in the shape function $F$, $\frac{\Delta F(\overrightarrow{k_1},\overrightarrow{k_2},\overrightarrow{k}_3)}{F(\overrightarrow{k_1},\overrightarrow{k_2},\overrightarrow{k_3})}$ to second order in $\e$,
due to Jacobson-Corely dispersion relation, is
\begin{equation}\label{DeltaFoverF}
\frac{\Delta F(\overrightarrow{k_1},\overrightarrow{k_2},\overrightarrow{k_3})}{F(\overrightarrow{k_1},\overrightarrow{k_2},\overrightarrow{k_3})}=(-5 + \frac{k_c^3}{k_t^3} - \frac{k_s^2}{2 k_t^2}) \e^2.
\end{equation}
\begin{figure}
\includegraphics[angle=0,scale=0.5]{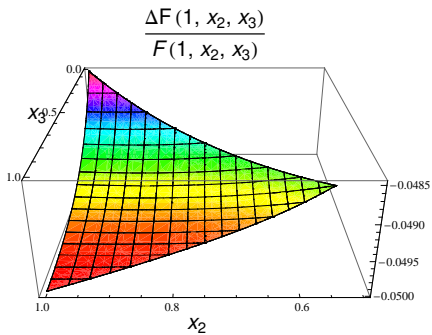}
\caption{The figure shows the dependence of $\frac{\Delta F(1,x_2,x_3)}{F(1,x_2,x_3)}$ on $x_2$ and $x_3$ setting  $\e=0.1$}\label{deltafoverF}
\end{figure}

We have plotted $\frac{\Delta
F(\overrightarrow{k_1},\overrightarrow{k_2},0.002)}{F(\overrightarrow{k_1},\overrightarrow{k_2},0.002)}$
in figure \ref{deltafoverF}, setting $\e=0.1$. As one can see from the
plot and the formula (\ref{DeltaFoverF}), the
general effect up to quartic order in the expansion is very mild and proportional to $\epsilon^2$. The modification is slightly
configuration-dependent and the equilateral configurations are more suppressed with respect to the local ones by a factor of $\sim$ one percent. One should also notice that the enfolded configurations are not enhanced with respect to the other ones.


\section{Explanation of the absence of enhancement factors in the
framework of particle creation} \label{secJCinterpretation}


We have seen in the previous section that no interference term seems
to appear in the bispectrum. This seems to contradict the expectation
from the fact that a modified dispersion relation can be understood in
terms of a ``vacuum'' state for the perturbations given by an excited
state.

Comprehending this point is
crucial in showing that there will not be any enhancement for the
enfolded configurations.
In particular, the analysis arriving at
this result is based on an expansion of the Whittacker function in (\ref{KummerM-BesselJ})
valid for small $z= i \epsilon k^2\eta^2$: as the integral in the
three-point function (\ref{bispectrum}) goes up to $\eta \to -\infty$,
we would like to understand in other terms the physical reason behind the absence
of the interference terms.

To make sure that it does not depend on the
particular expansions used, we have performed another analysis
without using any series expansion. Finally, we have also considered
the approximated solution found with {\it i}) the gluing method, {\it
ii}) the WKB approximation and compared the different approaches.

\subsubsection{Analysis with complete solution in the region
  $\omega_{\text{phys}} > H$}\label{alternativeJCcomplete}

In this section we are going to solve the equation exactly in the region
  $\omega_{\text{phys}} > H$. The solutions are more manageable than before and
need not be expanded. We make a change of variables: $\chi=-\sqrt{2\epsilon}k\eta$, so
that (\ref{eom-cosm-perturb}) becomes
\beq
\partial_{\chi}^2 u_k +({\chi^2\ov 4} + {1\ov 2\epsilon}-{2 \ov \chi^2})u_k = 0 \,.
\eeq
We see that for $\chi^2 > 4\epsilon \leftrightarrow k|\eta| > 2$, we can approximate the equation as
\beq
\partial_{\chi}^2u_k +({{\chi^2}\ov 4} + {1\ov 2\epsilon})u_k = 0 \,.
\eeq
The general solution to this differential equation is \cite{Martin:2002kt}
\beq
u_k(\chi) = c_1 E(a, \chi)+
c_2 E^{\ast}(a, \chi) \, , \qquad
a \equiv -{1 \ov 2\epsilon} \, ,
\eeq
where $E(\nu, y)$ are complex parabolic cylinder functions \cite{Abramowitz}.

The solution has two asymptotic regions, A and B, corresponding to
${\chi^2 \ov 4} < |a|$ and ${\chi^2 \ov 4} > |a|$. Using the property of parabolic cylinder
functions \cite{Abramowitz}, one finds in A
\beq
E(a, \chi) \underset{\chi^2 \gg |a|}{\sim}
{2^{{1 \ov 4}} \ov \epsilon^{{1 \ov 4}}k^{{1 \ov 2}}|\eta|^{{1 \ov 2}}}\,\,
(\sqrt{2\epsilon}k\eta)^{{i \ov \epsilon}}\,\,
e^{i {\epsilon \ov 2} \eta^2 k^2+i{\pi \ov 4}+i\phi_2} \, ,
\eeq
where $\phi_2\equiv \text{arg}(\Gamma({1 \ov 2}-{i \ov
2\epsilon}))$. We reproduce  the correct WKB asymptotic (see
\cite{Martin:2002kt}) choosing
\beq
c_1 = {e^{-i{\pi \ov 4}-i\phi_2} \ov 2^{{5 \ov 4}}\epsilon^{{1 \ov 4}}k^{{1 \ov 2}}},
\qquad c_2 = 0.
\eeq
In region B we use the asymptotic of $E(a, \chi)$ for $\chi \ll a$, obtaining
\beq
u_k \sim c_1 {1 \ov 2^{1 \ov 4}}
\left|{\Gamma({1 \ov 4}-i{1 \ov 4\epsilon}) \ov \Gamma({3 \ov 4}-i{1 \ov 4\epsilon})}\right|^{{1 \ov 2}}
e^{i{\pi \ov 4}} \, e^{-i k \eta + O(\epsilon^2\eta^2)}\, .
\eeq

We therefore find that no term proportional to $e^{i k\eta}$ appears
in the mode function, $u_k$. Since the Wightman function is
proportional to $u'^{\ast}_k$, we are assured that no interference
term will appear in the
Wightman function. We will now make a comparison with the result
obtained using the gluing and the WKB approximations.

\subsubsection{Analysis using gluing and WKB methods}\label{gluingmethod}

In this section we approximate the solution using the gluing
method. We also discuss the WKB approach. As for the gluing,
to start with we notice that
there are three different asymptotic regions for equation
(\ref{eom-cosm-perturb}) and its solution:
 \be \label{JCregions}
&(k|\eta|)^2 \, < \, 2  && \text{region I} \nonumber \\
& 2 \, < \, (k|\eta|)^2 \, < \, \epsilon^{-2} && \text{region II} \nonumber \\
& (k|\eta|)^2  \, > \, \epsilon^{-2} \qquad &&
\text{region III} \nonumber \\
\ee

One approximates the solution in
the various regions in (\ref{JCregions}) and asks for the continuity
of the mode functions and their first derivatives across the
boundaries of the regions. We start from region III, where,
being $k|\eta| > \e^{-1}$, equation (\ref{eom-cosm-perturb}) can be
approximated by \eqref{eom-approx}. The asymptotic solution in region
${\rm III}$ is given by \eqref{WKB},
where we have taken only the positive frequency mode.

In region ${\rm II}$ we have instead a solution of the form
\beq \label{planewavesII}
u_k = {\alpha_k \ov \sqrt{2k}} e^{-i k \eta}+
{\beta_k \ov \sqrt{2k}} e^{i k \eta} \, .
\eeq
The coefficients
$\alpha_k, \beta_k$ are fixed imposing the continuity conditions at
the time $\eta_c$ given by
\beq \label{etac}
 \eta_c= - \frac{1}{\e k} \, ,
\eeq
which separates regions {\rm II} and {\rm III}.
We find
\beq
\begin{cases}
\alpha_k =  e^{-{i \ov 2\epsilon}}\left(1+i{\epsilon \ov 4}\right) & \\
\beta_k = -i e^{i{3 \ov 2\epsilon}}\left({\epsilon \ov 4}\right)
\label{betacoefficient}
\end{cases} \, ,
\eeq
where we have used $k|\eta_c| = \epsilon^{-1}$.
We can now investigate the appearance of enhancements, expected
from the interference terms that would appear in
(\ref{bispectrum}) due to the form of the solution
(\ref{planewavesII}). In fact, the integrand of the
bispectrum given by (\ref{bispectrum}) is
proportional to the product of three Wightman functions. The integral
domain over $\eta$ can be divided into regions $\rm {I, II}$ and
${\rm III}$ where we can use the approximate solutions in each region. In
particular, because of the form of the solution in region {\rm II},
see (\ref{planewavesII}),
in parallel with the results
of \cite{Holman:2007na} one finds enhancement factors multiplying
the rest of the bispectrum from the
interference terms. However, from \eqref{betacoefficient} these
factors are of the order of
\beq
|\beta_k k \eta_c|\sim \epsilon |k\eta_c| = 1 .
\eeq
That is, there is
no {\em large} enhancement factor, even before
taking into account the suppression that comes from the 2D projection
on the CMB surface.

Furthermore, this method also overestimates the corrections to the power
spectrum. This can be seen by comparing to the WKB-method, where there is no
distinction between regions II and III. By matching the WKB solution,
valid there, with the growing and decaying modes solving the equation
of motion in region {\rm I}, one
readily finds that there are no
interference terms, and that the spectrum is corrected only at the order
$\epsilon^2$, in agreement with the result obtained with the exact
solution. Failure of gluing method in obtaining the correct amount of modification to the power spectrum was encountered in \cite{Joras:2008ck} too.

\vspace{0.4cm}

\paragraph{Contact with the NPS method and importance of the Wronskian
condition}

~

The NPHS procedure for modifying the vacuum in presence of
trans-Planckian physics makes direct contact with the result we have
found in this section via the gluing method, if we identify the cutoff
with the scale $p_c$. In that case, the evolution in region ${\rm III}$ is
accounted for by an excited state implementing boundary conditions for the
perturbations that lead to the conditions (\ref{betacoefficient}) at
the $k-$dependent junction time \eqref{etac}.

Also in the NPHS scenario where the Bogolyubov parameter is determined by
the minimization of the uncertainty relation
between the field and its conjugate momentum  and by the Wrosnkian
condition, there is actually no large enhancement factor. This
descends from the fact that in that case
\cite{Danielsson:2002kx}
\beq
|\beta_k|_{(NPHS)} = (2|k\eta_c|)^{-1} \sim {H \ov \Lambda}
= \epsilon
\eeq
where we have used the definition (\ref{etac}). Furthermore, in other
prescriptions for the choice of vacuum in line with the NPHS approach
\cite{Martin:2002kt}, the
coefficient $\beta_k$ is even further suppressed, as powers of
${H \ov \Lambda}$. Therefore, it seems
that in most of the emergent cases in the cutoff procedure the second
Bogolyubov factor is very constrained and will not lead to any large
enhancement factor.


\section{Discussion, outlooks and conclusion}\label{discussionconclusion}


Quantum gravity effects are expected to modify the standard Lorentzian
dispersion relation at very high energies, close to the Planck
scale. One notable example is Ho\v{r}ava-Lifshitz gravity
\cite{Horava:2009if,Horava:2009uw}, where  higher order derivative
corrections yield a modified dispersion relation. Such high energy
modifications could leave detectable signatures in the temperature fluctuations of
the CMBR, allowing an experimental detection of this aspect of
high energy theories.

Using the illustrative example of the Corley-Jacobson (JC) modified
dispersion relation with positive coefficient, we were able to
investigate the modification to the bispectrum in the case where there
is no violation of the WKB condition at early times, and to
compare these with the expectations deriving from the NPHS approach.
The field equations for the JC dispersion relation could be solved
exactly, so that the complete formula for the bispectrum, in series of
the ratio $\e = {H \over p_c}$, can be obtained. In this ratio, $H$ is
the Hubble parameter during inflation, while $p_c$ is the scale at
which the modification to the linear dispersion relation become
important.

It has been found that the leading correction to the standard bispectrum is
suppressed by a factor $\e^2$ and no configuration of momenta leads to
compensating factors that enhance the correction. The modification is
slightly more pronounced for the equilateral configurations by a
factor of one percent with respect to the local ones. The
analysis of the bispectrum in two different approximations and in
the framework of
particle creation leads to the conclusion that if the modified
dispersion relation does not violate the WKB condition at early times,
the particle production is too small to generate large
modifications to the bispectrum. In particular, using the WKB
approximation gives results in qualitative and quantitative agreement
with those obtained with the exact solution, while the gluing method
proposed by Brandenberger and Martin, which relates more directly with
the NPHS approach to trans-Planckian physics, is in partial disagreement.

One would expect more favorable possibilities when the modified
dispersion relations do violate the WKB condition at early
times. However, this point needs to be verified, as the real presence
of enhancements could depend strictly on the form of the dispersion
relations. We leave this to future investigation.

We have also shown that in most cases within the New Physics Hyper-surface
approach to modeling trans-Planckian physics, where at the scale $p_c$
of new physics the theory is cutoffed and a boundary condition is
imposed on the fields, there is no occurrence of large enhancement factors
in the bispectrum, not even for the enfolded configurations studied in
\cite{Holman:2007na}. Most notably, the enhancement is only of order
$O(1)$ when the Bogolyubov
coefficients accounting for particle creation are determined by
minimizing the uncertainty relation and by the Wronskian condition,
which yields the largest correction to the power spectrum.


\section*{Acknowledgments}


A.A. is supported by the G\"{o}ran Gustafsson Foundation. U.D. is
supported by the the G\"{o}ran Gustafsson Foundation and the Swedish
Research Council (VR).
Diego Chialva is supported by a Postdoctoral F.R.S.-F.N.R.S. research
fellowship via the Ulysses Incentive Grant for the Mobility in Science
(promoter at the Universit\'e de Mons: Per Sundell). A.A. acknowledges useful discussions with M. M. Sheikh-Jabbari and G. Shiu.


\end{document}